\RequirePackage{fix-cm}
\documentclass[twocolumn,epjc3]{svjour3}  
\smartqed  
\RequirePackage{graphicx}
\newcommand{\AC}{A_\mathrm{C}}
\newcommand{\AFB}{A_\mathrm{FB}}
\newcommand{\ttb}{{t\bar{t}}}
\newcommand{\pT}{p_{\mathrm{T}}}
\newcommand{\GeV}{\mathrm{GeV}}
\newcommand{\fb}{\mathrm{fb}}
\journalname{Eur. Phys. J. C}
\begin{document}

\title{The top quark charge asymmetry in $t\bar{t}\gamma$ production at the LHC}
\subtitle{}

\author{Jonas Bergner\thanksref{e1,addr1}
        \and
        Markus Schulze\thanksref{e2,addr1}
}

\thankstext{e1}{e-mail: Jonas.Bergner@physik.hu-berlin.de}
\thankstext{e2}{e-mail: Markus.Schulze@physik.hu-berlin.de}

\institute{Institut f\"ur Physik, Humboldt-Universit\"at zu Berlin, D-12489 Berlin, Germany \label{addr1}
}


\maketitle

\begin{abstract}
We consider the top quark charge asymmetry in the process $pp \to t\bar{t}+\gamma$ at the 13 TeV LHC.
The genuine tree level asymmetry in the $q\bar{q}$ channel is large with about $-12\%$.
However, the symmetric $gg$ channel, photon radiation off top quark decay products,
and higher order corrections wash out the asymmetry and obscure its observability.
In this work, we investigate these effects at next-to-leading order QCD and check the robustness of theoretical predictions. 
We find a sizable perturbative correction and discuss its origins and implications. 
We also study dedicated cuts for enhancing the asymmetry and show that a measurement 
is possible with an integrated luminosity of $150\,\fb^{-1}$.

\end{abstract}

\section{Introduction}
\label{sec:intro}

The charge asymmetry in the fermion annihilation process $f\bar{f} \to f' \bar{f}'$ is a well-studied phenomenon of Quantum Electrodynamics (QED)~\cite{Berends:1973fd,Berends:1982dy} 
and Quantum Chromodynamics (QCD)~\cite{Nason:1989zy,Beenakker:1990maa}.
Even though $\mathrm{C}$ (charge) and $\mathrm{P}$ (parity) are good symmetries of QED and QCD, there can be interference between charge-odd amplitudes that
lead to asymmetric terms under $p_{f'} \leftrightarrow p_{\bar{f}'}$.
For $2\to2$ kinematics the $\mathrm{C}$-odd interference appears at next-to-leading order (NLO) for the first time, causing a suppression by one power of the coupling
constant and therefore yields a numerically small asymmetry. 

At particle colliders where the initial state is not charge symmetric (e.g. at $e^+ e^-$ or $p\bar{p}$ colliders), a non-vanishing charge asymmetry $\AC$
translates into a $\mathrm{P}$-violating forward\- backward asymmetry $\AFB$. 
This feature received a lot of attention in the case of top quark pair production at the Tevatron. 
The NLO QCD theory prediction of $\AFB^{\ttb} \approx 5\%$~\cite{Kuhn:1998jr} was in long lasting tension with the experimentally measured values 
by CDF and DZero, which were about two standard deviations higher~\cite{Peters:2012ji}.
The dust settled after more data was collected and NNLO QCD and NLO electroweak corrections~\cite{Hollik:2011ps,Bernreuther:2012sx,Czakon:2014xsa} were accounted for in the theory calculations.
Now, the best prediction yields $\AFB^\ttb=9.5\pm 0.7\%$ \cite{Czakon:2014xsa}, which has to be compared to $\AFB^\ttb=10.6\pm 3\%$ from DZero~\cite{Abazov:2014cca} and $\AFB^\ttb=16.4\pm 5\%$ from CDF~\cite{Aaltonen:2012it}.

Recently, the top quark charge asymmetry enjoyed a revival at the Large Hadron Collider (LHC) because the delicate interference effects can be used as a sensitive probe for new physics searches, 
see e.g. Refs.~\cite{Rodrigo:2010gm,Kuhn:2011ri,Han:2012qu,Ko:2012ud,Hagiwara:2012gy} and references in~\cite{Aguilar-Saavedra:2014kpa}. 
The charge symmetric initial state at the LHC does, however, not produce a forward-backward asymmetry and makes the effect harder to capture. 
An observable effect can still be obtained thanks to the different parton distributions of valence and sea quarks in the proton, which, in conjunction with a non-zero $\AC$ cause
anti-top quarks to be scattered more centrally than top quarks. 
The canonical definition of the charge asymmetry at the LHC is
\begin{eqnarray} \label{eq:AC}
     & \AC & = \frac{\sigma^\mathrm{asymm.}}{\sigma^\mathrm{symm.}}
      \quad\mathrm{with}
\\
      &\sigma^\mathrm{asymm.} &= \sigma(\Delta y > 0)-\sigma(\Delta y < 0),  
\nonumber  \\
      &\sigma^\mathrm{symm.}&=\sigma(\Delta y > 0)+\sigma(\Delta y < 0),
\nonumber
\end{eqnarray}
where $\Delta y = |y_t| - |y_{\bar{t}}|$ is the difference of the absolute top and anti-top quark rapidities.
For $pp\to\ttb$ at $\sqrt{s}=8$~TeV, the best prediction $\AC^\ttb= 0.9\%$~\cite{Czakon:2017lgo} is in agreement with current experimental measurements~\cite{Aad:2015noh,Khachatryan:2015oga}
that are, however, also compatible with zero given the smallness of the effect.

An interesting twist enters the discussion when studying top quark pair production in association with massless gauge bosons. 
Hadronic production of $\ttb+\mathrm{jet}$ is one example that has been studied extensively~\cite{Dittmaier:2008uj,Melnikov:2010iu,Berge:2012rc,Alte:2014toa}.
$\mathrm{C}$ asymmetric interference terms enter already at leading order causing a sizable negative value of the asymmetry. 
Somewhat surprisingly, the inclusion of higher order corrections shifts the LO value by more than 100\%~\cite{Dittmaier:2008uj} into the positive direction. 
This feature appears in contrast to $\ttb$ production where the asymmetry at NLO does not receive large corrections at NNLO~\cite{Czakon:2014xsa}. 
In Ref.~\cite{Melnikov:2010iu} a reasoning for this pattern was given based on a separation of {\it soft} and {\it hard} degrees of freedom
that enter the asymmetry at different orders of perturbation theory. 

In this work, we investigate the charge asymmetry for top quark pair production in association with a photon at the LHC.
An experimental measurement of this quantity was not undertaken at the Tevatron and is not yet achieved at the LHC. 
This is surprising since $\ttb+\gamma$ production is particularly interesting for probing physics beyond the Standard Model.
For example, in a pioneering study the authors of Ref.~\cite{Aguilar-Saavedra:2014vta} investigated the use of $\AC^{\ttb\gamma}$ to resolve cancellation mechanisms between 
up-type and down-type initial states arising from possible new physics contamination of the SM signal. 
Here, we elevate previous studies at LO to NLO QCD precision.
This is motivated by a foreseeable measurement in the near future and by the importance higher order corrections played in the similar process $pp \to\ttb+\mathrm{jet}$.
A first NLO QCD calculation for $\AC$ in $\ttb+\gamma$ production was presented in Ref.~\cite{Maltoni:2015ena} for stable top quarks. 
In this work put special attention to a realistic description of the process, accounting for the full decay chain in the lepton+jets final state, 
$b \bar{b} \ell \nu  j j+\gamma$, including all spin correlations, photon emission off all charged particles, and NLO QCD corrections in production and decay.
We demonstrate that every of these features is crucial for a reliable description of the charge asymmetry in this process. 
Moreover, we devise dedicated selection cuts to enhance the asymmetry while simultaneously maintaining the statistical significance of a measurement.

\section{Setup}
\label{sec:setup}
We consider the process $pp \to \ttb+\gamma \to b \bar{b} \ell \nu  j j+\gamma$ at $\sqrt{s}=13$~TeV,
summing over $\ell=e^+,e^-,\mu^-$ and $\mu^+$.
In this final state, the top quark momenta can be reconstructed unambiguously from the decay products since the neutrino momentum is constrained 
by momentum conservation. 
Hence, the top quark rapidities in Eq.~(\ref{eq:AC}) can be calculated unambiguously.
Top quarks are treated in the narrow-width approximation (NWA). 
We require intermediate on-shell states which result from a $q_t^2$-integration 
over their (undistorted) Breit-Wigner propagator,
\begin{equation}
   \int \!\! \mathrm{d}  q_t^2
   \left| \frac{1}{(q_t^2-m_t^2 + \mathrm{i} \Gamma_t m_t)} \right|^2 
\!\to\! \frac{\pi }{m_t \Gamma_t} \int \!\! \mathrm{d}  q_t^2
    \delta(q_t^2-m_t^2)
\end{equation}
in the limit $\Gamma_t/m_t \to 0$.
It is well known that this treatment leads to a parametric approximation of the cross section up to terms $\mathcal{O}(\Gamma_t  \big/ m_t)$.
In this case, the amplitude for $\ttb$ production factorizes according to 
\begin{eqnarray} \label{eq:PrDk}
   \mathcal{M}^\mathrm{NWA}_{ij\to t\bar{t}\to b \bar{b} f\bar{f}f'\bar{f}'}
   \;=\;
   \mathcal{P}_{ij \to t\bar{t}} \otimes \mathcal{D}_{t\to b f\bar{f}}
   \otimes \mathcal{D}_{\bar{t}\to \bar{b} f'\bar{f}'}~,
\end{eqnarray}
where $\mathcal{P}_{ij \to t\bar{t}}$ describes the $t\bar{t}$ production process and $\mathcal{D}_{t\to b f\bar{f}}$ the top quark decay dynamics.
The symbol $\otimes$ indicates the inclusion of spin correlations.

The factorization for the $\ttb+\gamma$ process is obtained from Eq.~(\ref{eq:PrDk}) by inserting a photon in either of the three terms, unfolding it into a sum of three terms
at $\mathcal{O}(\alpha)$. 
As a consequence, $\ttb+\gamma$ production is governed by two very different dynamics:
(i) photons can be emitted in the hard scattering process of $\ttb$ production, followed by the top decays; and 
(ii) photons can be emitted off the top quark decay products, which is preceded by $\ttb$ production. 
We refer to these two mechanisms as {\it photon radiation in production} and {\it radiative top quark decays}, respectively.  
An equivalent way of phrasing this circumstance is: The photon can be radiated either before 
or after the top quarks went on-shell.

\begin{table*}
\caption{$\ttb+\gamma$ cross sections, the charge asymmetry and the corresponding significance $\mathcal{S}$ for different sets of cuts. 
Scale variation uncertainties for the cross sections are given in relative terms ($\pm$) and in absolute numbers (brackets) for the asymmetry and significance.
}
\begin{center}
\label{tab:1}     
\begin{tabular}{c|c|c|c}
     & cuts Eq.~(\ref{eq:acc_cuts}) & cuts Eq.~(\ref{eq:acc_cuts})+(\ref{eq:supp_cuts}) & cuts Eq.~(\ref{eq:acc_cuts})+(\ref{eq:supp_cuts})+$|y_\gamma|>1.0$  \\
\hline     
$\sigma^\mathrm{symm.}_\mathrm{LO}$        & $837\,\fb \pm 25\%$     & $301\,\fb \pm 28\%$      & $126\,\fb \pm 25\%$ \\[2ex]
$\sigma^\mathrm{asymm.}_\mathrm{LO}$       & $-10.6\,\fb \pm 21\%$   & $-9.2\,\fb \pm 22\%$     & $-5.6\,\fb \pm 21\%$ \\[1ex]
\hline
$\sigma^\mathrm{symm.}_\mathrm{NLO}$       & $1708\,\fb \pm 19\%$    & $647\,\fb \pm 21\%$      & $287\,\fb \pm 22\%$ \\[2ex]
$\sigma^\mathrm{asymm.}_\mathrm{NLO}$      & $-7.8\,\fb \pm 6\%$     & $-6.4\,\fb \pm 6\%$      & $-4.8\,\fb \pm 2\%$\\[1ex]
\hline
$\AC^\mathrm{NLO}$                         & $-0.5(1)\%  $           &  $-1.0(2)\%  $           & $-1.7(4)\%  $ \\[1ex]
$\mathcal{S}_\mathrm{NLO}$                 & $2.3(3)\sigma$          &  $3.1(4)\sigma$          & $3.5(4)\sigma$      
\end{tabular}
\end{center}
\end{table*}

To account for a finite detector volume and resolution we require 
\begin{eqnarray} \label{eq:acc_cuts}
    &\pT^\gamma \ge 20\,\GeV,
    \quad
    |y_\gamma| \le 2.5,
    \quad
    R_{\gamma \ell} \ge 0.2,
    \quad
    R_{\gamma \mathrm{jet}} \ge 0.2,
\nonumber   
\\
    &\pT^\ell \ge 15\,\GeV,
    \quad
    |y_\ell| \le 5.0,
    \quad
    \pT^\mathrm{miss} \ge 20\,\GeV,
\nonumber
\\
    &\pT^\mathrm{jet} \ge 15\,\GeV,
    \quad
    |y_\mathrm{jet}| \le 5.0.
\end{eqnarray}
We define jets by the anti-$k_\mathrm{T}$ jet algorithm~\cite{Cacciari:2008gp} with $R=0.3$ and request at least two $b$-jets. 
Photons in a hadronic environment are defined through the smooth-cone isolation~\cite{Frixione:1998jh} with $R=0.2$.
We perform our calculations within the TOPAZ framework described in Ref.~\cite{Melnikov:2011ta}. 
The input parameters to our calculation are 
\begin{eqnarray}
   & \alpha=1/137, 
   \quad
   G_\mathrm{F} = 1.16639 \times 10^{-5} \,  \GeV^{-2},
\nonumber   
\\
   &m_t = 173\, \GeV,
   \quad
   M_W = 80.419\, \GeV,
\end{eqnarray}
from which follows
\begin{eqnarray}
   &\Gamma_t^\mathrm{LO}=1.495\,\GeV,
    \quad
    \Gamma_t^\mathrm{NLO}=1.367\, \GeV,
\nonumber   
\\
    &\Gamma_W^\mathrm{LO}=2.048\, \GeV,
    \quad
    \Gamma_W^\mathrm{NLO}=2.118\, \GeV
\end{eqnarray}
at $\mu_\mathrm{R}=m_t$.
We use the parton distribution functions NNPDF31\_nlo\_ as\_0118\_luxqed~\cite{Bertone:2017bme}, 
with the corresponding running of the strong coupling constant $\alpha_s$.

\section{Higher Order Corrections to $\AC$}
\label{sec:corrections}

In the following we discuss the impact of higher order QCD corrections to the charge asymmetry and the 
cross section of $pp \to \ttb+\gamma$.
Applying the cuts in Eq.~(\ref{eq:acc_cuts}) and setting renormalization and factorization scales 
to $\mu_0=m_t$, we find
\begin{eqnarray}
\AC^\mathrm{LO} = -1.3\%,
\quad
\AC^\mathrm{NLO} = -0.5\%.
\end{eqnarray}
The leading order value of $-1.3\%$ arises from the 
asymmetric contribution in $q\bar{q} \to \ttb+\gamma$, which has a genuine asymmetry of $-12\%$ that  
is diluted by the symmetric $gg$ channel and radiative decays.
Comparing LO and NLO asymmetries, the relative shift by more than $-60\%$ is striking
and might raise questions about the perturbative convergence of this quantity.
We therefore investigate the different perturbative corrections in greater detail
and note that $\AC$ by itself is not an observable.
Only $\sigma^\mathrm{asymm.}$ and $\sigma^\mathrm{symm.}$, i.e. the numerator and denominator of $\AC$ are experimentally accessible quantities. 
The term $\sigma^\mathrm{symm.}$ is just the total cross section, which is known to receive large corrections (see e.g. Refs.~\cite{Melnikov:2011ta,PengFei:2009ph}). 
We confirm this feature within our setup and find the leading and next-to-leading order cross sections 
\begin{eqnarray}
\label{eq:sigmasymm}
\sigma_\mathrm{LO} &=& 837\,\fb \pm 25\%,
\nonumber \\
\sigma_\mathrm{NLO} &=& 1708\,\fb \pm 19\%.
\end{eqnarray}
Renormalization and factorization scales are varied by a factor of two around the central scale $\mu_0$
and the respective cross sections are symmetrized. 
The large perturbative correction of $104\%$ and
the marginal reduction of scale uncertainty is a combination of various effects:
Firstly, the dominant $gg$ channel receives a sizable ($\approx \! +80\%$) perturbative correction.
The main contribution arises from tree level type $\ttb\gamma+g$ configurations, where the gluon constitutes a hard resolved jet (similar features were observed in Ref.\cite{Melnikov:2011ta}). 
Secondly, the kinematics of the light jets from $W\to jj$ are significantly restricted at leading order because of jet cuts and the jet algorithm. 
This restriction is lifted  when an additional jet is allowed at next-to-leading order. 
It affects all partonic channels and leads to yet another increase of the NLO cross section by about $+20\%$. 
While the size of this kinematic effect cannot be estimated with scale variation, we believe it is 
sufficiently saturated at NLO, yielding a realistic and reliable prediction. 
Lastly, the $qg$ initial state enters at NLO for the first time and is responsible for the 
sizable residual scale dependence of the NLO cross section in Eq.~(\ref{eq:sigmasymm}).

\begin{figure*}
  \includegraphics[width=0.5\textwidth]{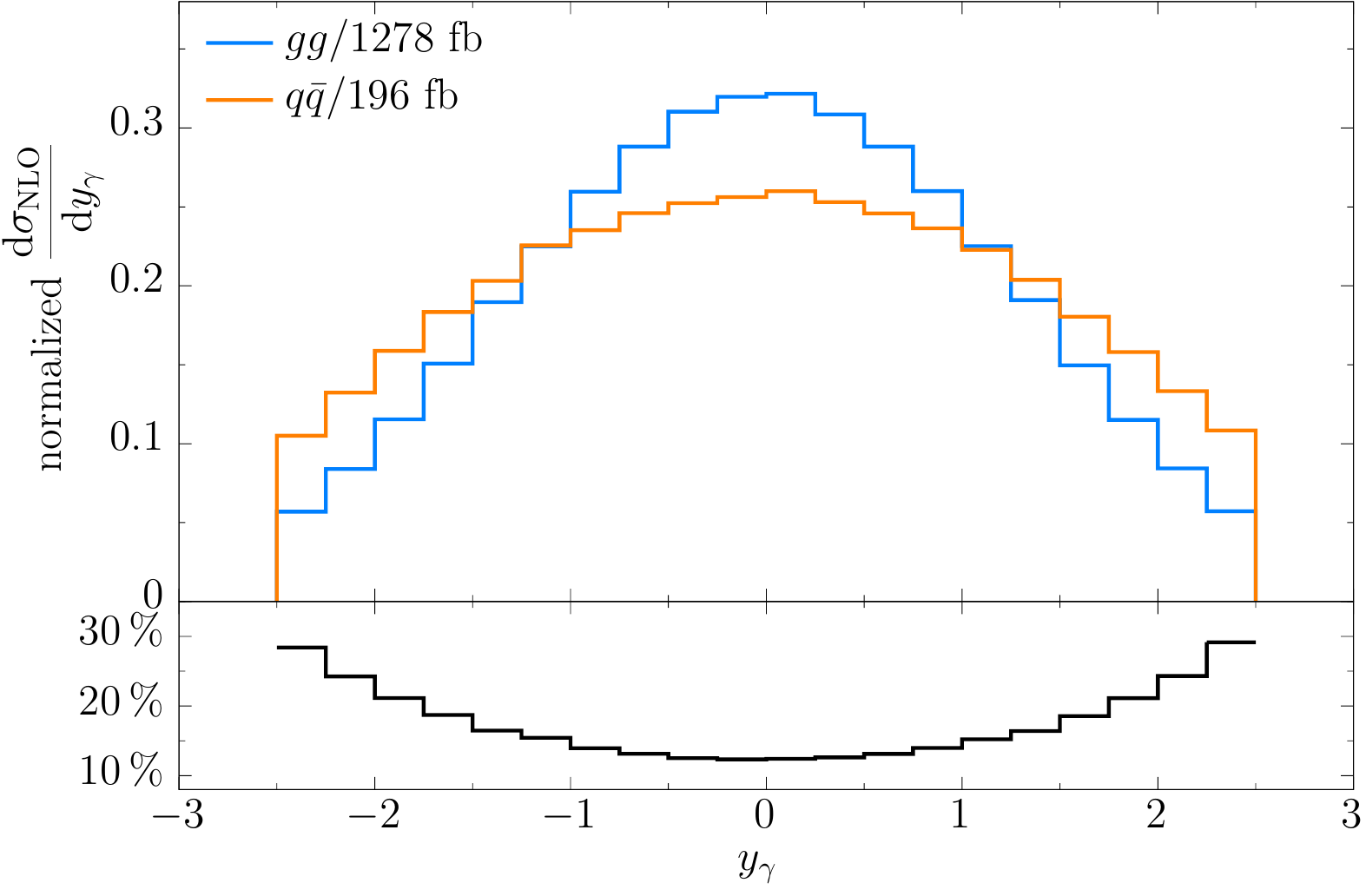}
  \includegraphics[width=0.5\textwidth]{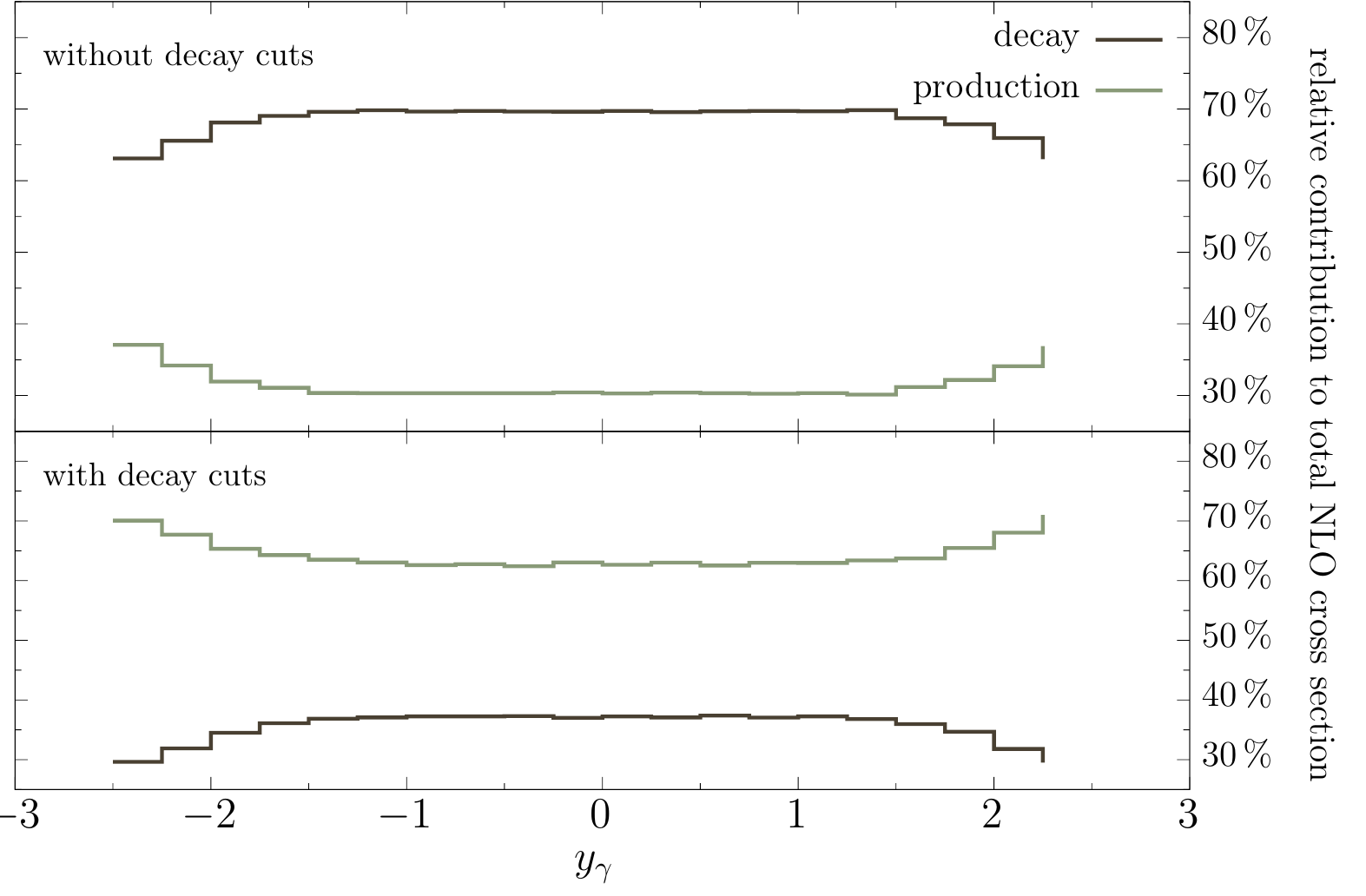}  
\caption{Left: Normalized photon rapidity distribution of the $gg$ and $q\bar{q}$ initial states (upper pane) 
and relative contribution of the $q\bar{q}$ initial state with respect to the total contribution (lower pane).
Right: Relative contribution of photon radiation in production and photon radiation in decay with respect to the total contribution
as a function of the photon rapidity. The upper pane includes the selection cuts in Eq.~(\ref{eq:acc_cuts}). The lower pane 
includes the suppression cuts in Eq.~(\ref{eq:supp_cuts}) in addition to Eq.~(\ref{eq:acc_cuts}).
}
\label{fig1}      
\end{figure*}

Let us now discuss the perturbative correction of the numerator of the asymmetry. 
We find a very different behavior 
\begin{eqnarray}
\label{eq:sigasymsym}
\sigma^\mathrm{asymm.}_\mathrm{LO} &=& -10.7\,\fb \pm 20\%,
\nonumber \\
\sigma^\mathrm{asymm.}_\mathrm{NLO} &=& -8.0\,\fb \pm 7\%.
\end{eqnarray}
In contrast to $\sigma^\mathrm{symm.}$ in Eq.~(\ref{eq:sigmasymm}), the asymmetric piece receives a 
moderate $-25\%$ correction and enjoys a significantly reduced scale dependence. 
Hence, the perturbative convergence seems under good control. 
This conclusion is further supported by another observation. 
Adopting the reasoning of Ref.~\cite{Melnikov:2010iu} for $\ttb+\mathrm{jet}$ production, the asymmetry is governed by {\it soft} and {\it hard}
degrees of freedom, which enter at different stages of perturbation theory. 
In the limit where the cross section is dominated by logarithms of $p_\mathrm{T,cut}^{\gamma} \big/ m_t$
one finds \cite{Melnikov:2010iu}
\begin{eqnarray}
\label{eq:Amechanism}
A_{q\bar{q}\to\ttb\gamma}^\mathrm{NLO} \approx A_{q\bar{q}\to\ttb\gamma}^\mathrm{LO} + A_{q\bar{q}\to\ttb}^\mathrm{NLO}.
\end{eqnarray}
The {\it soft} degrees of freedom are contained in $A_{q\bar{q}\to\ttb\gamma}^\mathrm{LO}$ because it is 
generated dominantly by a soft photon exchange. 
Beyond LO, new asymmetric contributions appear from {\it hard} exchanges that are related to the asymmetry in $\ttb$ production $A_{q\bar{q}\to\ttb}^\mathrm{NLO}$.
To study these dynamics for our case, we perform an independent NLO QCD calculation for $pp\to\ttb\to b \bar{b} \ell \nu  j j$ at $\sqrt{s}=13$~TeV, using the 
same cuts as in Eq.~(\ref{eq:acc_cuts}).
We find $A_{q\bar{q}\to\ttb}^\mathrm{NLO}=+2.9\%$. 
Together with $A_{q\bar{q}\to\ttb\gamma}^\mathrm{LO}=-12.0\%$ and $A_{q\bar{q}\to\ttb\gamma}^\mathrm{NLO}=-8.9\%$,
this nicely supports the prediction in Eq.~(\ref{eq:Amechanism})\footnote{Note that in contrast to Eq.~(\ref{eq:sigasymsym}), only photon radiation in the production is considered here.}.
Consequently, we follow the arguments presented in Ref.~\cite{Melnikov:2010iu} and suggest that even higher order corrections (i.e. beyond NLO QCD) should stabilize the prediction of 
\\
$\sigma^\mathrm{asymm.}$  
and will not drastically shift its value. 

\begin{figure*}
\begin{center}
  \includegraphics[width=0.40\textwidth]{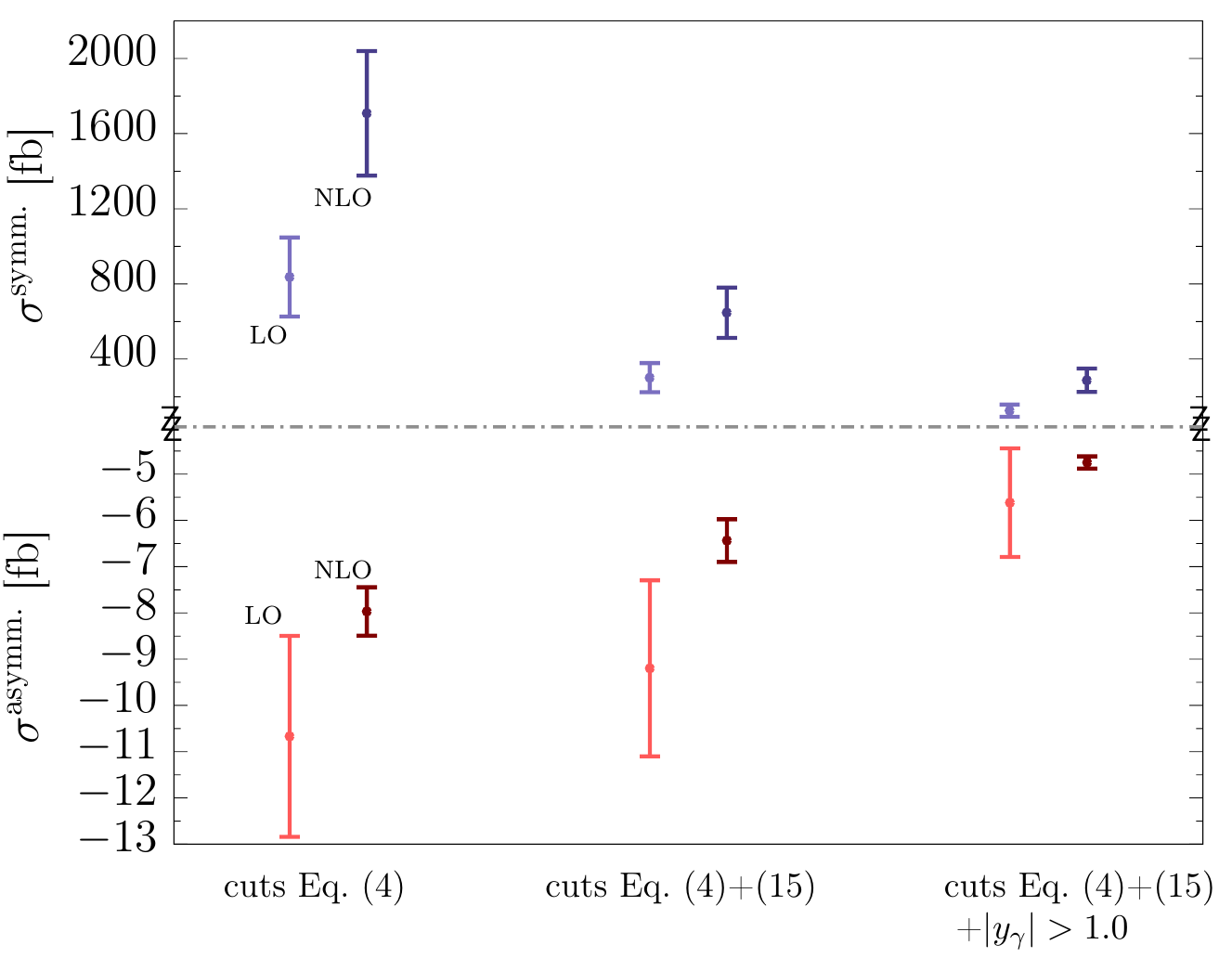}
  \includegraphics[width=0.59\textwidth]{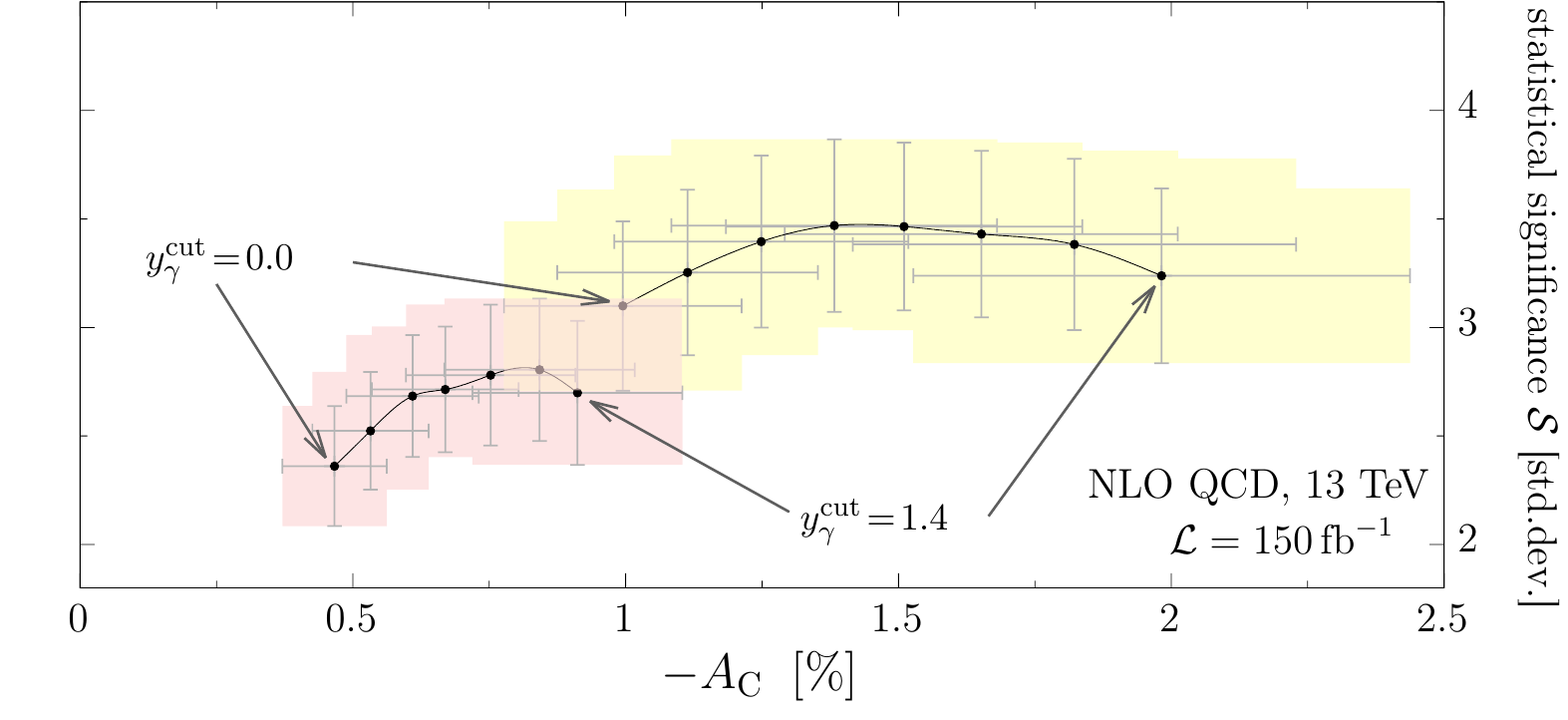}
\caption{Left: Symmetric and asymmetric cross sections at LO and NLO QCD for different sets of cuts. The error bands correspond to a
scale variation by a factor two around the central scale $\mu_0$.
Right: Statistical significance  as a function of the (negative) charge asymmetry at NLO QCD. 
The lower line corresponds to cuts Eq.~(\ref{eq:acc_cuts}) and the upper line arises from the additional application of cuts Eq.~(\ref{eq:supp_cuts}).
Each line consists of seven data points for increasing cut $|y_\gamma| \ge 0.0,\, 0.2,...,1.2,1.4$.
}
\label{fig2}   
\end{center}
\end{figure*}

From these studies we conclude that the genuinely asymmetric cross section of $\ttb+\gamma$ production is perturbatively under good control, 
whereas symmetric contributions from $gg$ and $qg$ initial states are converging slower with sizable scale dependence.
The resulting relative uncertainty for the asymmetry  
\begin{equation}
\label{eq:deltaAC}
\frac{\delta\AC}{\AC} = \sqrt{ \left(\frac{\delta\sigma^\mathrm{asymm.}}{\sigma^\mathrm{asymm.}}\right)^2 
                             + \left(\frac{\delta\sigma^\mathrm{ symm.}}{\sigma^\mathrm{ symm.}}\right)^2 }
\end{equation}
is therefore dominated by $\delta\sigma^\mathrm{symm.}\big/\sigma^\mathrm{symm.}=\pm 19\%$.
The significance of a measurement (assuming statistical uncertainties only) is
\begin{eqnarray}
   \mathcal{S} &=& |\AC| \big/ \delta N
   \quad \mathrm{with} \quad 
   \delta N = 1 \big/ \sqrt{N}
\nonumber \\
   &=& |\AC| \, \sqrt{\mathcal{L} \times \sigma_\mathrm{NLO}}.
\end{eqnarray}
The corresponding numerical values can be found in the first column of Table~\ref{tab:1} 
and are illustrated in the first column of Fig.~\ref{fig2} (left) for an 
integrated luminosity of $150\,\fb^{-1}$.

\section{Analysis and Results}
\label{sec:results}

We proceed with a study of dedicated phase space cuts to enhance the charge asymmetry. 
The basic idea is to isolate asymmetric contributions while suppressing symmetric ones. 
The dominant asymmetric contribution originates from quark anti-quark annihilation with photon 
radiation in the production, $q\bar{q} \to \ttb + \gamma$.
Large symmetric contributions arise from $gg$ scattering and photon emission in the top quark decay stage.

A suppression of the $gg$ channel over the $q\bar{q}$ channel is notoriously difficult to achieve. 
However, we find that one can use shape differences in the photon rapidity distributions to separate the two channels.
Fig.~\ref{fig1} (left, upper pane) illustrates this feature for the normalized distributions. The lower pane
shows the relative percentage of $q\bar{q}$ versus $gg$ as a function of a photon rapidity at  NLO QCD.
It is evident that an increasing lower cut value  
\begin{equation}
\label{eq:yphcut}
 |y_\gamma| \ge y_\gamma^\mathrm{cut}
\end{equation}
 is enhancing this relative percentage, while, at the same time, reducing the overall cross section. 

The feature of radiative top quark decays is the second source of large symmetric contributions. 
Splitting the cross section into photon radiation in production (prod) and radiative decays (dec), we find for 
the cuts in Eq.~(\ref{eq:acc_cuts}) 
\begin{equation}
     \sigma_\mathrm{NLO}^{\ttb+\gamma}= 526\,\fb\,(\mathrm{prod}) + 1182\,\fb\,(\mathrm{dec}) = 1708\,\fb.
\end{equation}
Almost $70\,\%$ of the total rate is due to $\ttb$ production followed by a radiative top quark decay.
This is a somewhat counter-intuitive picture as one typically imagines the $\ttb+\gamma$ final state as being 
produced altogether in the hard collision. 
We suppress radiative top quark decays using invariant masses of the decay products.
To start, we associate the two $b$-jets with the {\it correct} side of the decay chain 
($b$-jets belong to the $t$ decay chain, $\bar{b}$-jets belong to the $\bar{t}$ decay chain).
This is achieved by pairing $b$-jet and leptonic decay chain which minimize $\{ m_{\ell b_1},m_{\ell b_2}\}$.
The other $b$-jet is associated with the hadronic decay chain.
Subsequently, we consider the minima
\begin{eqnarray}  \label{eq:supp_cuts}
     \min_{x \in D_i \cup D_{i\gamma}} \bigg\{ m_x^2 - m_t^2 \bigg\}, \quad i=\mathrm{\ell,h}
\end{eqnarray}
for $D_\ell=\{ b\ell\nu, b\ell\nu j \}$, $D_{\ell \gamma}=\{ b\ell\nu\gamma, b\ell\nu j\gamma,\}$,
$D_\mathrm{h}=\{ b j j, b j j j \}$, and $D_{\mathrm{h} \gamma}=\{ b j j\gamma, b j j j \gamma\}$.
If kinematics is such that $x \in D_{\ell\gamma} $ or $x \in D_{\mathrm{h}\gamma} $, we consider it a radiative top quark decay event and reject it. 
All other events are kept. 
We find that these selection criteria are robust under QCD corrections, and we 
believe that the impact of off-shell effects is small because a smearing of the invariant masses around the top quark Breit-Wigner peak will not significantly change the minimization procedure. 
This assertion can, in principle, be checked thanks to the off-shell calculation presented in Ref.~\cite{Bevilacqua:2018woc}.
Fig.~\ref{fig1} (right) shows the relative contribution of photon emission in the production and radiative top quark decays.
The upper pane shows the two contributions without the cuts of Eq.~(\ref{eq:supp_cuts}), the lower pane shows the contributions when the cuts are included. 
It is evident that this procedure works very efficiently in selecting photon emission in production. 
Moreover, the rapidity distribution is flat and remains flat after the cuts. 
Hence, the cuts for $gg$ suppression (Eq.~(\ref{eq:yphcut})) and radiative decays do not interfere with each other. 
\\

In the following, we study the charge asymmetry as a function of the cuts in Eq.~(\ref{eq:yphcut}) and Eq.~(\ref{eq:supp_cuts}) including NLO QCD corrections.
It is evident that applying the cuts on the one hand increases the asymmetry, and on the other hand reduces the cross section, 
therefore lowering the statistical significance of a measurement. 
Hence, we try to optimize the cuts such that the two competing effects are balanced. 
We vary the lower photon rapidity cut $y_\gamma^\mathrm{cut}$ from $0.0$ to $1.4$ in steps of $0.2$.
The results are given in the second and third column of Table~\ref{tab:1} and displayed in Fig.~\ref{fig2}.
These are the main results of this work. 
We find that the perturbative pattern that we discussed in Sect.~\ref{sec:corrections} persist
if the cuts Eq.~(\ref{eq:yphcut}) and Eq.~(\ref{eq:supp_cuts}) are added.
The asymmetric contribution receives a moderate NLO correction with small scale dependence.  
In contrast, the symmetric cross section gets large corrections and exhibits $\approx 20\%$ scale dependence. 
This uncertainty feeds into the uncertainty of the asymmetry $\AC$ as the dominating one.

From Table~\ref{tab:1} it is evident that the additional cuts significantly enhance the asymmetry. 
The initial value of $-0.5\%$ is doubled when the radiative decay suppression cuts in Eq.~(\ref{eq:yphcut}) are applied. 
Further, it is more than tripled when $|y_\gamma|>1$ is required in addition.
The relative uncertainties remain roughly constant at about $20\%$.
The statistical significance is boosted to values above $3\sigma$ for an integrated luminosity of $\mathcal{L}=150\,\fb^{-1}$.

Fig.~\ref{fig2} (right) illustrates the dependence on the cuts in more detail and allows to find the optimal cut values. 
We plot the significance $\mathcal{S}$ over the negative asymmetry $\AC$.
The connected dots show the dependence on the monotonically increasing value $y_\gamma^\mathrm{cut}$,
with cuts in Eq.~(\ref{eq:acc_cuts}) (lower dotted line) and cuts in Eqs.~(\ref{eq:acc_cuts})+(\ref{eq:supp_cuts}) (upper dotted line). 
The colored bands indicate the corresponding uncertainties obtained from Eq.~(\ref{eq:deltaAC}) for $\AC$ and similar for $\mathcal{S}$.
Comparing the pink and yellow bands it is obvious that the radiative decay suppression is very effective in 
enhancing the asymmetry and the statistical significance.
Yet the uncertainties are somewhat inflated. 
Following the dotted lines of increasing $y_\gamma^\mathrm{cut}$, we observe that the asymmetry can be 
strongly enhanced while the significance receives a mild increase and later deteriorates for too large values. 
The optimal point appears at $y_\gamma^\mathrm{cut} \approx 1.0$.

\section{Summary}
\label{sec:summary}

We study the top quark charge asymmetry in the lepton+jet final state of $\ttb+\gamma$ production at the 13~TeV LHC. 
The asymmetry is an Abelian effect of interference between diagrams of even and odd charge-parity, 
a phenomenon that is well-studied for $\ttb$ production at the Tevatron and the LHC. 
The $pp \to \ttb+\gamma$ process is interesting because it exhibits an asymmetry already at leading order, 
which is significantly larger than in $\ttb$ production.
We present perturbative corrections to this observable, including top quark decays, and discuss uncertainties and 
enhancement strategies. 
We find that the asymmetric cross section is converging well and is under good theoretical control. 
In contrast, the symmetric cross section receives sizable corrections. 
As a result, leading order predictions turn out to be unreliable and next-to-leading order predictions carry sizable uncertainties. 
Yet, we find arguments to support the reliability of our NLO results within their uncertainties. 
In addition, we present a set of tailored cuts for enhancing the asymmetry by more than a factor of three
such that a measurement with $150\,\fb^{-1}$ should be possible at the LHC.

\begin{acknowledgements}
We thank Ivor Fleck, Manfred Kraus, Till Martini, and Peter Uwer for fruitful feedback and discussions.
We are grateful for computing resources provided by AG PEP. 
\end{acknowledgements}

\bibliographystyle{spphys}       
\bibliography{acbib}   

\end{document}